\documentclass[preprint]{revtex4}

\usepackage{graphicx}
\usepackage{dcolumn}
\usepackage{bm}
\usepackage{hyperref}

\begin{document}

\preprint{preprint Physical Review B}

\title{Magnetic resonance as an orbital state probe}

\author{A.~A.~Mozhegorov}
\email{alexey.mozhegrov@usu.ru}
\author{A.~V.~Larin, A.~E.~Nikiforov, L.~E.~Gontchar, A.~V.~Efremov}
\affiliation{
Ural State University,\\
Lenin ave. 51, 620083, Ekaterinburg, Russia}%

\date{\today}

\begin{abstract}
Magnetic resonance in ordered state is shown to be the direct
method for distinguishing the orbital ground state. The example of
perovskite titanates, particularly, LaTiO$_3$ and YTiO$_3$, is
considered. External magnetic field resonance spectra of these
crystals reveal glaring qualitative dependence on assumed orbital
state of the compounds: orbital liquid or static orbital structure.
Theoretical basis for using the method as an orbital state probe
is grounded.
\end{abstract}


\maketitle

Recently, a lot of spin and orbital phases and their phase
transitions attract intent investigation due to interplay of these
degrees of freedom, particularly in transition-metal (TM)
oxides.~\cite{ImadaReview1998} Fundamental physical properties
revealed in such systems are still a subject for a discussion.
Among the phenomena which attract the most attention there is a
superexchange interactions driven rich spin-orbital quantum phase
diagram proposed to exist in perovskite titanates and
vanadates.~{\cite{Khaliullin2000,Khaliullin2003,Oles2005}}

Whether orbital liquid state present in real compound or not ---
is the question which have given rise to hot debates especially
concerning the simpler system --- $R\text{TiO}_3$, $R$ is
rare-earth element or Y.

\begin{figure}
\includegraphics[width=6cm,keepaspectratio]{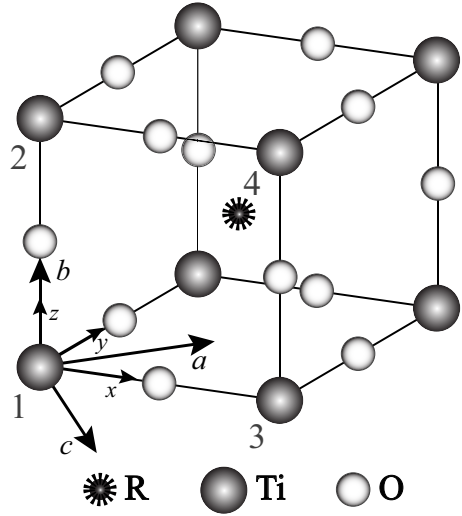}
\caption{\label{fig-structure} $R$TiO$_3$ crystal structure.
\emph{Pnma} ({\textbf a}, {\textbf b}, {\textbf c}) and
pseudocubic ({\textbf x}, {\textbf y}, {\textbf z}) axes are
shown. Numbers denote Ti sublattices.}
\end{figure}

In wide temperature range for different $R$, titanates are known
to possess orthorhombic crystal structure
~\cite{Komarek2007,Cwik2003} (which is often called
''quasi-cubic'') with \emph{Pnma} space group (see
Fig.~\ref{fig-structure}). GdFeO$_3$-type distortions
($T_{1g}$-distortions), which are present in these crystals, are
believed to control magnetic structure and properties of the
compounds through the influence on their orbital ground
state.~\cite{ImadaReview1998,MochizukiImadaPRL,MochizukiImadaNJP}

The discussion of titanates orbital ground state was started by
Khaliullin and Maekawa~\cite{Khaliullin2000} who proposed the
superexchange model with dynamical quenching of local orbital
moments in a simple cubic lattice of Ti$^{3+}$ ions with no static
orbital order but with fixed magnetic arrangement --- in contrast
to usual Goodenough-Kanamori picture. This approach based on the
Kugel-Khomskii model~\cite{KugKhomUFN1982} perfectly explains the
unusual reduction of Ti$^{3+}$ spin~\cite{Meijer1999,Ulrich2002}
and anomalous isotropic spin-wave spectrum~\cite{Keimer2000} found
experimentally, but contradicts to NMR~\cite{KiyamaItoh2003} and
XAS~\cite{Haverkort2005} experiments as well as some
crystal-field~\cite{MochizukiImadaPRL,MochizukiImadaNJP,SchmitzLa,SchmitzY}
and density-functional~\cite{Solovyev2006,Pchelkina2005}
calculations. However, it is supported by recent Raman scattering
experiments~\cite{Ulrich2006} revealing orbital excitations and
thus making some orbital fluctuations possible.

In this context La and Y titanates are believed to be of special
interest for investigators as these two ions stand at the opposite
ends of rare-earths and Y series with different ionic radii.
$T_{1g}$-distortions in YTiO$_3$ with ''smaller'' Y are much
greater of these in LaTiO$_3$, which has the biggest radius of
$R$-ion in the whole series. One more feature which makes these
two crystals more interesting is their magnetic ground states:
lanthanum titanate is antiferromagnetic~\cite{Cwik2003} with
strong isotropic superexchange \emph{J} of about $15.5$
meV,~\cite{Keimer2000} whereas yttrium titanate is ferromagnetic
with $J\approx-2.75$~meV.~\cite{Ulrich2002} Moreover, almost
isotropic superexchange couplings both in LaTiO$_3$ and its sister
compound are in contrast with the situation in manganites and
cuprates which have pronounced superexchange anisotropy.

Keeping in mind a huge collection of all above mentioned facts,
one still can not find a compromise between them and can not
answer what ground state is actually realized in titanates. That
indicates a puzzling situation: on the one hand the system is
already studied in many aspects, on the other hand some of its
properties are not clear yet. To meet this challenge, the definite
model of a titan perovskite oxide is developed in the current
paper on the particular example of Y and La compounds

We start from the crystal-field Hamiltonian with explicit
electron-lattice interaction, that is vibronic Hamiltonian (VH):
\begin{widetext}
\begin{eqnarray}
H_{vib} & = & H_{lin}+H_{QQ}+H_{R}=\nonumber\\
        & = &
            \left[
            V_e
                \left(Q_{\theta}X_{\theta}+Q_{\epsilon}X_{\epsilon}\right)+
            V_t
                \left(Q_{\xi}X_{\xi}+Q_{\eta}X_{\eta}+Q_{\zeta}X_{\zeta}\right)
\right]+\nonumber\\
        & + &
            \left[\begin{array}{l}
            V_a
                \left(Q^2_x+Q^2_y+Q^2_z\right)X_{A1}+
            V_b
                \left\{
                    \left(2Q^2_z-Q^2_x-Q^2_y\right)X_{\theta}+
                    \sqrt{3}\left(Q^2_x-Q^2_y\right)X_{\epsilon}
                \right\}+\\
            V_c
                \left(Q_yQ_zX_{\xi}+Q_xQ_zX_{\eta}+Q_xQ_yX_{\zeta}\right)]
            \end{array}\right]+\nonumber\\
        & + &
            \left[\begin{array}{l}
            {V_e}^R
                \left(Q^R_{\theta}X_{\theta}+Q^R_{\epsilon}X_{\epsilon}\right)+\\
            {V_t}^R
                \left\{
                    \left(Q^R_{\xi,1}+Q^R_{\xi,2}\right)X_{\xi}+
                    \left(Q^R_{\eta,1}+Q^R_{\eta,2}\right)X_{\eta}+
                    \left(Q^R_{\zeta,1}+Q^R_{\zeta,2}\right)X_{\zeta}
                \right\}
            \end{array}\right]~.
\label{Hvib}
\end{eqnarray}
\end{widetext}
Here $Q_\Gamma$ and $Q^R_\Gamma$
($\Gamma=\theta,\epsilon,\xi,\eta,\zeta,x,y,z$) are symmetrized
shifts of oxygen and $R$-ions which are nearest and next-nearest
Ti$^{3+}$ neighbours correspondingly. These shifts are obtained
from accurate crystal structure data for LaTiO$_3$~\cite{Cwik2003}
and YTiO$_3$.~\cite{Komarek2007} $X_\Gamma$ are symmetric orbital
operators, acting on the 3d-$t_{2g}$ triplet, and $V_\alpha$
($\alpha=e,t,a,b,c$) and $V^R_\alpha$ ($\alpha=e,t$) are
electron-lattice coupling constants.~\cite{Nikiforov1980} The
first two square brackets, which are $H_{lin}$ and $H_{QQ}$,
represent Ti$^{3+}$ 3d-$t_{2g}$ electron interactions with nearest
oxygen linear and quadratic symmetric shifts correspondingly,
whereas $H_R$ is responsible for this electron and $R$-ions shifts
coupling.

This Hamiltonian with ab initio $V_\alpha$ (calculated with GAMESS
package~\cite{GAMESS1,GAMESS2}) gives ground state with static
orbital structure of the form predicted by Mochizuki and
Imada,~\cite{MochizukiImadaPRL,MochizukiImadaNJP} that is the
following orbital functions (3d-$t_{2g}$ cubic basis set):
$\psi_{1,2}(\text{La})=\psi_{3,4}(\text{La})\approx1/\sqrt{3}\left(-\xi-\eta\pm\zeta\right)$,
which is almost trigonal, in LaTiO$_3$ and
$\psi_{1,2}(\text{Y})\approx1/\sqrt{2}\left(\mp\xi+\zeta\right)$,
$\psi_{3,4}(\text{Y})\approx1/\sqrt{2}\left(\mp\eta+\zeta\right)$
--- in YTiO$_3$. Here sign alternation reads sites 1(2) -- the upper
sign, and 3(4) -- the lower one.

We have irrefutable argument for reproducing results of previous
LDA-based ~\cite{Haverkort2005,Solovyev2006,Pchelkina2005} or
point-charges
~\cite{MochizukiImadaPRL,MochizukiImadaNJP,SchmitzLa,SchmitzY}
investigations by using (\ref{Hvib}). All these studies either
were based on oversimplified model~\cite{MochizukiImadaPRL,MochizukiImadaNJP} (giving unlikely
structure parameters) or did not reveal the mechanisms of the
ground state formation.~\cite{Haverkort2005,SchmitzLa,SchmitzY,Solovyev2006,Pchelkina2005}
The present calculation based on (\ref{Hvib}) showed that $H_R$
plays crucial role in the formation of solitary orbital singlet
with its segregation at $\approx0.20$~eV in LaTiO$_3$ and
$\approx0.15$~eV in YTiO$_3$. A half of these gaps is produced by the
$R$-ion crystal field. The influence of the remainder of the
crystal on Ti$^{3+}$ orbital state is negligible.

Using the low-energy spectrum obtained from (\ref{Hvib}) we then exploit common Kugel-Khomskii (KK) method within
the Hubbard model.~\cite{KugKhomUFN1982} Thus, arriving to isotropic superexchange, one then can perform
Moriya's approach~\cite{Moriya1960} for treating antisymmetric terms~\cite{Nikiforov1971} of the effective
$S\text{-}\frac12$ spin-Hamiltonian (ESH) introduced below:
\begin{equation}
H_{ef\!f}=J_{ij}\left({\bf S}_i\cdot{\bf S}_j\right)+
{\bf D}_{ij}\left({\bf S}_i\times{\bf S}_i\right)+
S^\alpha_iA^{\alpha\beta}_{ij}S^\beta_j+
H^\alpha_ig^{\alpha\beta}_iS^\beta_i,
\label{ESH}
\end{equation}
where $J_{ij}$ stands for isotropic superexchange between $i$-th and $j$-th magnetic ions,
${\bf D}_{ij}$ is Dzyaloshinskiy-Moriya vector, ${\textsf A}_{ij}$ is symmetric anisotropy tensors,
${\textsf g}_i$ is g-factor and $\bf H$ represents external magnetic field.

\begin{figure}
\includegraphics[width=8cm,keepaspectratio]{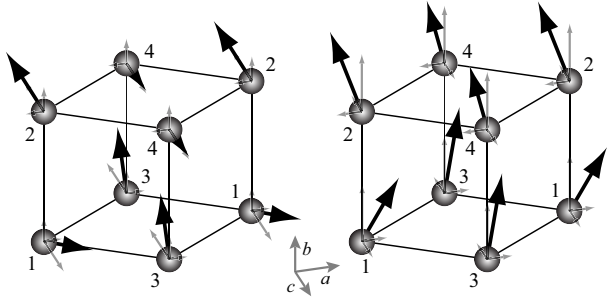}
\caption{\label{fig-MagStructure}
Schematic magnetic structures of LaTiO$_3$ (up) and YTiO$_3$ (bottom).
Gray arrows denote magnetization projections (\emph{Pnma}).}
\end{figure}

The Hubbard model parameters: energy of electron hopping from the
$m$-th orbital on one site to the $n$-th on the neighboring one in
strictly cubic system $t_{mn}$, on-site Coulumb repulsion of a
pair of electrons $\text U$ and intra-atomic electronic exchange
interaction ${\text J}_{\text H}$ --- are taken from LDA-based
calculation.~\cite{Solovyev2006} For LaTiO$_3$
    $t_{mm}\approx0.25$~eV,
    $t_{m \neq n}\approx0.12$~eV,
    ${\text U}\approx3.20$~eV,
    ${\text J}_{\text H}\approx0.61$~eV.
For YTiO$_3$
    $t_{mm}\approx0.27$~eV,
    $t_{m \neq n}\approx0.08$~eV,
    ${\text U}\approx3.45$~eV,
    ${\text J}_{\text H}\approx0.62$~eV.
Indexes in $t_{mn}$ denote 3d-$t_{2g}$ orbitals, namely $\xi$,
$\eta$, $\zeta$. All interactions in ESH, which follow from these
parameters, are listed in Tables~\ref{table1},~\ref{table2}.

\begin{table}
\caption{\label{table1} Single crystal g-factors of LaTiO$_3$ and
YTiO$_3$. The \emph{Pnma} coordinate system is used.}
\begin{ruledtabular}
\begin{tabular}{ccc}
$R$ & La & Y
\\ \hline
 ${\textsf g}_i$
    & $\left(
        \begin{array}{ccc}
         1.93 &  0.03 & -0.05 \\
         0.03 &  1.89 & -0.04 \\
        -0.05 & -0.04 &  1.93
        \end{array}
        \right)$
    & $\left(
        \begin{array}{ccc}
         1.93 & -0.04 & -0.04 \\
        -0.04 &  1.87 &  0.02 \\
        -0.04 &  0.02 &  1.96
        \end{array}
        \right)$
\end{tabular}
\end{ruledtabular}
\end{table}

This calculation of superexchange couplings, although it is not something new, is to be reproduced
because of extreme sensitivity of these parameters to the orbital state. This well-known feature
of the KK treatment should be considered as (in comparison with previous studies
~\cite{MochizukiImadaPRL,MochizukiImadaNJP,Haverkort2005,SchmitzLa,SchmitzY,Solovyev2006,Pchelkina2005,Eremina2004})
we have obtained new
$\psi_i(\text{La})$ and $\psi_i(\text{Y})$ , $i=1 \ldots 4$.

\begin{table*}
\caption{\label{table2} Magnetic interactions (meV) in LaTiO$_3$
and YTiO$_3$. The \emph{Pnma} notations are used.}
\begin{ruledtabular}
\begin{tabular}{ccccccc}
 $R$ & $J_{12}$ & $J_{12}$ & ${\bf D}_{12}$ & ${\bf D}_{13}$ & ${\textsf A}_{12}$ & ${\textsf A}_{13}$
 \\ \hline
 La
    & 13.21
    & 16.12
    & $\left(
        \begin{array}{c}
         0.58 \\
         0 \\
        -2.80
        \end{array}
        \right)$
    & $\left(
        \begin{array}{c}
         0 \\
        -0.36 \\
        -0.17
        \end{array}
        \right)$
    & $\left(
        \begin{array}{ccc}
        -0.07 &  0    &  0.06 \\
         0    & -0.08 &  0 \\
         0.06 &  0    & -0.34
        \end{array}
        \right)$
    & $\left(
        \begin{array}{ccc}
        -0.02 &  0    &  0 \\
         0    & -0.03 & -0.02 \\
         0    & -0.02 & -0.01
        \end{array}
        \right)$
 \\
 Y
    & -2.77
    & -2.72
    & $\left(
        \begin{array}{c}
         0.60 \\
         0 \\
        -0.21
        \end{array}
        \right)$
    & $\left(
        \begin{array}{c}
         0 \\
        -0.38 \\
        -0.11
        \end{array}
        \right)$
    & $\left(
        \begin{array}{ccc}
        -0.21 &  0    &  0.06 \\
         0    & -0.03 &  0 \\
         0.06 &  0    & -0.06
        \end{array}
        \right)$
    & $\left(
        \begin{array}{ccc}
        -0.69 &  0    &  0 \\
         0    & -0.10 & -0.09 \\
         0    & -0.09 & -0.16
        \end{array}
        \right)$
\end{tabular}
\end{ruledtabular}
\end{table*}

By using the Hamiltonian~(\ref{ESH}) one can obtain magnetic
ground state as well as magnetic excitations in both compounds.
The magnetic structure type for both crystals is $(A_x,F_y,G_z)$
with major $G$-component in LaTiO$_3$ and $F$ --- in YTiO$_3$ (see
Fig.~\ref{fig-MagStructure}) --- in excellent agreement with
neutron scattering experiments.~\cite{Meijer1999,Ulrich2002} Here
we should emphasize two special features in $J_{ij}$ which are
crucial for obtaining correct magnetic structure. First is
considering Hund's coupling, which is ${\text J}_{\text H}/{\text
U}$ smaller then ''common'' superexchange,~\cite{Anderson1959}
proportional to $t^2_{mn}/{\text U}$ for titanates. The second is
explicit introduction of the Ti--O--Ti bond angle ($\varphi$)
--dependence of $t_{mn}$.~\cite{Moskvin1977} These two give for a
pair of Ti$^{3+}$ ions along the {\textbf b} axis (\emph{Pnma}):
\begin{equation}
J_{12}=a
    \left(b-\frac{{\text J}_{\text H}}{\text U}\right)-
    c
    \left(d+\frac{{\text J}_{\text H}}{\text U}\right)
    {\sin}^2\varphi,
\label{Jz}
\end{equation}
where $a=3.54(2.61)$, $b=6.37(1.66)$, $c=43.30(14.70)$ and $d=0.84(0.92)$ are numeric coefficients for
La(Y) compound. These coefficients depend on the particular Ti$^{3+}$ orbital state, on $t_{mn}$ and $\text U$.
Ti--O--Ti bond angle $\varphi$ for lanthanum titanate is about $153^\circ$ and $142^\circ$ for YTiO$_3$. Without
${\text J}_{\text H}$ one can not obtain experimentally observed magnetic structure in both compounds
simultaneously as Schmitz \emph{et al.}~\cite{SchmitzLa,SchmitzY} couldn't. And in strictly cubic system
(without ${\sin}^2\varphi$--dependence of $J_{ij}$) one is not able to obtain the correct orbital
structure as Khaliullin and Maekawa couldn't~\cite{Khaliullin2000} for LaTiO$_3$.

It is important to mention that both anisotropic terms of the ESH, namely Dzyaloshinskiy-Moriya
interaction and symmetric anisotropy, should be considered, otherwise the static magnetic order couldn't
exist.~\cite{Shekhtman1992} This is not always kept by investigators.~\cite{Ulrich2002,Keimer2000}

Actually, the above result is not unique as its different parts were obtained by several authors.
~\cite{MochizukiImadaPRL,MochizukiImadaNJP,SchmitzLa,SchmitzY,Solovyev2006,Pchelkina2005,Eremina2004}
It is reproduced to illustrate the realistic model with reasonable parameters. But this model reveals
particular mechanisms of titanates orbital and magnetic ground state formation that was not performed before.

Now we turn to the core idea of the paper. That is drastic dependence of the particular kind of magnetic
excitations on orbital ground state.

\begin{figure}
\includegraphics[width=7cm,keepaspectratio]{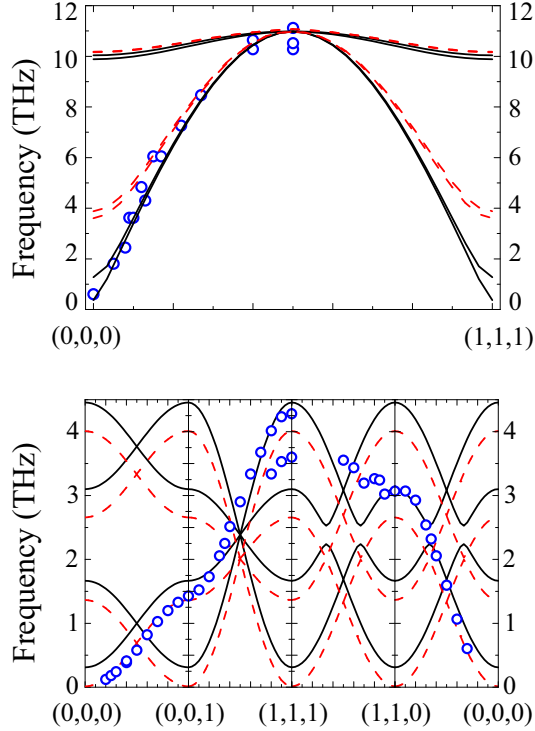}
\caption{\label{fig-SW}
(Color online)
Spin wave dispersions for LaTiO$_3$ (up) and YTiO$_3$ (bottom).
Circles denote experimental data, taken from
Ref.~\onlinecite{Keimer2000} for LaTiO$_3$ and from
Ref.~\onlinecite{Ulrich2002} for YTiO$_3$. Simulated curves are
black for static orbital order and dashed red for orbital liquid.}
\end{figure}

\begin{figure}
\includegraphics[width=7cm,keepaspectratio]{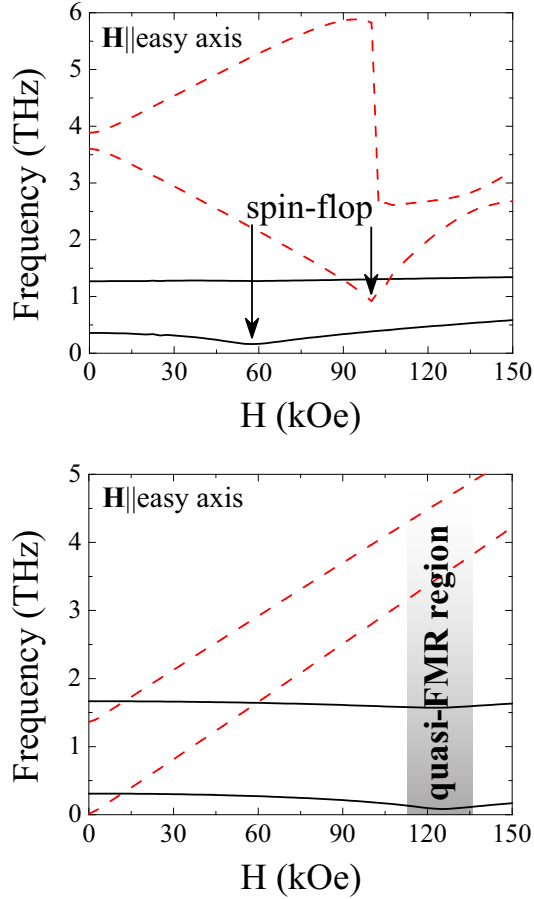}
\caption{\label{fig-MR}
(Color online)
Field dependencies of AFMR frequencies in LaTiO$_3$ (up) and YTiO$_3$ (bottom).
$\bf H$ is parallel to {\textbf c} for LaTiO$_3$ and {\textbf b} for YTiO$_3$ (\emph{Pnma}).
These directions are close to easy axes.
Curves are black for static orbital order and dashed red for orbital liquid.}
\end{figure}

We consider magnetic excitations, namely spin waves (SW) and antiferromagnetic/ferromagnetic resonance (AFMR/FMR)
spectra, in LaTiO$_3$ and YTiO$_3$ exploiting linear approximation. Simulations
(see Figs.~\ref{fig-SW},~\ref{fig-MR}) turned out to be surprising when comparing the real static orbital structure
with hypothetic orbital liquid state. The latter was simulated by averaging the superexchange parameters
through low-energy part of the spectrum (\ref{Hvib}).

There is an unexpected feature: spin-wave spectra exhibit almost no changes in two different orbital
states for both crystals. Little discrepancy in those simulations can be easily removed by the slight
fitting of the Hubbard model parameters, which fitting is really possible within these parameters calculation
discrepancies in different approaches.~\cite{Solovyev2006,Pchelkina2005,Solovyev2007} At the same time
there is drastic change in AFMR/FMR spectra for La and Y compounds. Orbital liquid shows up here
in two ways: rising anisotropy in LaTiO$_3$ and suppressing it in YTiO$_3$. This produces handbook curves
of AFMR and FMR field spectra, but with different anisotropic behavior.

This result is due to killing electronic distribution anisotropy factor in forming the magnetic
interactions by the orbital liquid. Thus, the lattice becomes the only source of magnetic anisotropy.
That is why the latter appears to be different in the compounds under consideration --- quite expectable
for La and Y crystals.

Unfortunately, no any attention was paid to such powerful and sensitive method of magnetic structure
and magnetic couplings investigation as antiferromagnetic/ferromagnetic resonance so far. The only
attempt to observe electron spin resonance (ESR) below the magnetic transition temperatures (this is AFMR
or FMR) and above them (EPR) was made by S. Okubo \emph{et al.}~\cite{Okubo1998} But, firstly, the AFMR signals
for LaTiO$_3$ were not obtained at all, that might be because of frequency limits of the equipment used.
And, secondly, this investigation was made only for powder samples and thus all direction-dependent effects
were wiped out. We can't but mention that if powder samples are used one neither will observe such an
interesting field spectra as represented in Fig.~\ref{fig-MR}, nor the strong orbital state
dependence of these spectra can be obtained.

Finally, we argue that up-to-day, magnetic resonance is the ultimate method for distinguishing between
static orbital order and orbital liquid. This method is a referee between opposite orbital states.
Particularly, AFMR/FMR experiments with single crystals should put a dot at the end of the discussion
what is the real orbital ground state in titanates --- the remarkable system with strong entanglement of
lattice, spin and orbital degrees of freedom.

We thank I.V. Solovyev for fruitful discussions. This work was
partially supported by the ''Dynasty'' foundation, by the CRDF REC-005,
by the Russian Federal Scientific Program ''The development of scientific potential of the high school''
and by the RFBR grant no.~08-02-00200.


\end{document}